\documentclass[aps,eqsecnum,preprint,preprintnumbers,12pt,amsfonts]{revtex4}
\usepackage{epsfig,psfrag}

\def\bea{\begin{eqnarray}}
\def\eea{\end{eqnarray}}
\def\nn{\nonumber}
\def\phic{\Phi_{\rm cl}}
\def\phib{\bar{\Phi}_{\rm cl}}  

\begin{document}

\title{Beyond the thin-wall approximation : precise numerical computation of prefactors in false vacuum decay}
\author{Gerald V. Dunne}\email{dunne@phys.uconn.edu}
\affiliation{Department of Physics, University of Connecticut, Storrs, CT 06269, USA}
\author{Hyunsoo Min} \email{hsmin@dirac.uos.ac.kr}
\affiliation{Department of Physics, University of Seoul, Seoul 130-743, Korea}

\begin{abstract}
We present a general numerical method for computing precisely the false vacuum decay rate, including the prefactor due to quantum fluctuations about the classical bounce solution, in a self-interacting scalar field theory modeling the process of nucleation in four dimensional spacetime. This technique does not rely on the thin-wall approximation. The method is based on the Gelfand-Yaglom approach to determinants of differential operators, suitably extended to higher dimensions using angular momentum cutoff regularization.  A related approach has been discussed recently by Baacke and Lavrelashvili, but we implement the regularization and renormalization in a different manner, and compare directly with analytic computations made in the thin-wall approximation. We also derive a simple new formula for the zero mode contribution to the fluctuation prefactor, expressed entirely in terms of the asymptotic behavior of the classical bounce solution.

\end{abstract}

\maketitle

\section{Introduction}
\label{intro}

The phenomenon of nucleation drives first order phase transitions in many applications in physics, most notably in particle physics, condensed matter physics, quantum field theory and cosmology. The semiclassical analysis of the rate of such a nucleation process was pioneered by Langer \cite{langer}, who identified a semiclassical saddle point solution that gives the dominant exponential contribution to the rate, with a prefactor to the exponential given by the quantum fluctuations about this classical solution. The nucleation rate is given by the quantum mechanical rate of decay of a metastable "false" vacuum, $\phi_-$, into the "true" vacuum, $\phi_+$. Decay proceeds by the nucleation of expanding bubbles of true vacuum within the metastable false vacuum \cite{langer,kobzarev,stone,coleman,affleck,voloshin1,weinberg1,voloshin-erice}. This picture may also be extended to finite temperature field theory \cite{linde,affleck2}. Computing the semiclassical prefactor requires the computation of the determinant of the differential operator associated with quantum fluctuations about the classical solution. This is a technically difficult problem. Thus, it is not possible to compute the fluctuation prefactor analytically, and so various approximation techniques have been developed. The most widely-studied is the so-called "thin-wall" approximation, in which the bubble wall thickness is small compared to the bubble radius \cite{stone,coleman,voloshin1}. This provides an elegant physical picture of nucleation and in this limit certain parts of the decay rate computation may be done analytically. A limited amount is also known analytically concerning the expansion beyond the leading thin-wall limit, in two dimensions \cite{konoplich,voloshin1,kiselev},  three dimensions \cite{munster,voloshin2+1}, and four dimensions \cite{kr}. Other important approaches to metastable decay use direct lattice simulations \cite{alford},  the average effective action \cite{strumia}, and phase shift techniques \cite{moss}. 

The point of this present paper is to reduce the calculation of the decay rate to a straightforward numerical computation, without relying on any such approximation or expansion. The starting point for our approach is an extremely elegant and simple method, due to Gelfand and Yaglom \cite{gy}, for computing the determinant of a one-dimensional differential operator. This technique is ideal for numerical implementation. However, its naive generalization to higher dimensions is divergent \cite{forman}. This is because in higher dimensions renormalization is important, and so we must regulate and renormalize the determinant. We present an analytic method of doing this using an angular momentum cutoff, and apply it to the problem of false vacuum decay. (This method has previously been used to compute the quark mass dependence of the fermion determinant for quarks in the presence of an instanton background \cite{dunne}.)
A related approach to false vacuum decay, also based on the Gelfand-Yaglom formula, has been developed recently by Baacke and Lavrelashvili \cite{baacke}, and we comment in Section \ref{ren} on the similarities and differences between our approaches.

\begin{figure}[h]
\includegraphics{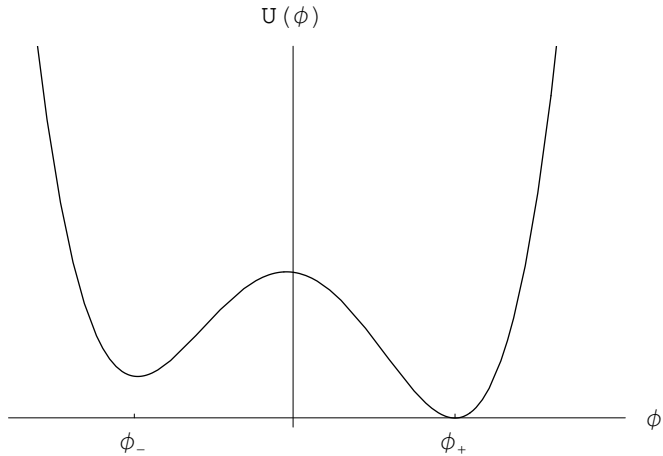}
\caption{Field potential $U(\phi)$ showing the true and false vacua, $\phi_+$ and $\phi_-$, respectively.}
\label{fig1}
\end{figure}

Consider the Euclidean classical action
\bea
S_{\rm cl}[\phi]=\int d^4 x\left( \frac{1}{2}(\partial_\mu \phi)^2+U(\phi)\right)\quad ,
\label{action}
\eea
where the potential $U(\phi)$ has two nondegenerate classical minima, $\phi_-  < \phi_+$, with $U(\phi_-) > U(\phi_+)$. For quantitative computations, and comparison with previous work, we consider the standard quartic potential \cite{kobzarev,coleman,konoplich,kiselev}, whose form is illustrated in Figure \ref{fig1}:
\bea
U(\phi)=\frac{\lambda}{8}\left(\phi^2-a^2\right)^2-\frac{\epsilon}{2 a}\left(\phi-a\right)\quad .
\label{pot1}
\eea
The parameter $\epsilon$ represents a constant external source breaking the degeneracy of the double well potential.
In four dimensional spacetime, the mass dimensions of the couplings are: $[\lambda]=0$, $[a]=1$, and $[\epsilon]=4$. We choose $a>0$, $\lambda>0$ and $\epsilon>0$, in which case the  two minima are $\phi_\pm=\pm a(1\pm \frac{\epsilon}{2\lambda a^4}+\dots ).$ Note that
\bea
U\left(\phi_-\right)-U\left(\phi_+\right)=\epsilon \left[1+O\left( \frac{\epsilon}{\lambda a^4}\right)\right]
\quad ,
\label{eps}
\eea
so for $\epsilon\ll \lambda a^4$, we see that  $\epsilon$ has the physical interpretation as the potential energy difference between the two classical vacua. This small $\epsilon$ limit is known as the ``thin-wall'' limit \cite{coleman} because in this case the bubbles of true vacuum within the false vacuum have thin walls compared to their radius.

\begin{figure}[h]
\includegraphics[scale=0.75]{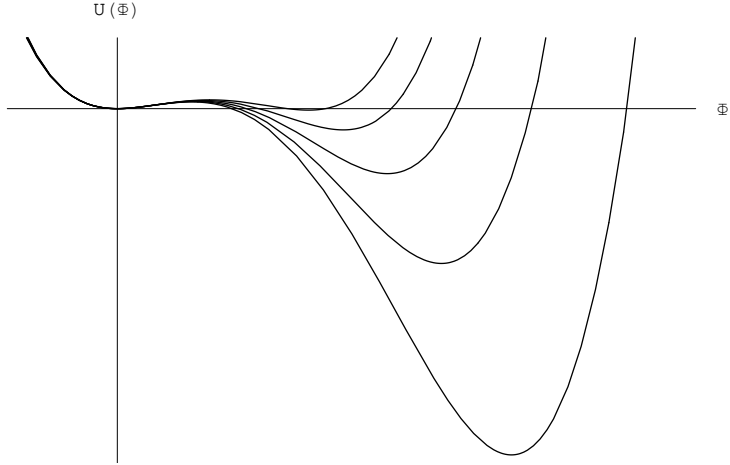}
\caption{Plots of the rescaled potential, $U(\Phi)=\frac{1}{2}\Phi^2-\frac{1}{2}\Phi^3+\frac{\alpha}{8}\Phi^4$, for  $\alpha=0.6, 0.7, 0.8 , 0.9, 0.99.$ As $\alpha$ approaches 1, the vacua become degenerate.}
\label{fig2}
\end{figure}

Expanding the field $\phi$ about the false vacuum $\phi_-$
\bea
\phi=\phi_- +\varphi\quad ,
\label{expansion}
\eea
and keeping terms up to dimension four, we find the potential, which is often considered directly in the literature \cite{baacke}
\bea
U(\varphi)=\frac{m^2}{2}\varphi^2-\eta \varphi^3 +\frac{\lambda}{8}\varphi^4\quad .
\label{pot2}
\eea
Here $m^2$ and $\eta$ are related to the original couplings by
\bea
m^2=\frac{\lambda}{2}\left(3\phi_-^2-a^2\right) \quad ; \quad \eta=\frac{\lambda}{2}|\phi_- | \quad .
\label{meta}
\eea
In order to describe semiclassical tunneling, it is useful to rescale the field $\varphi$ and the spacetime coordinates as 
\bea
\bar{x}= m x\quad ;  \quad\varphi =\frac{m^2}{2\eta} \Phi
\label{rescale}
\eea
Then the classical action in terms of these dimensionless quantities is :
\bea
S_{\rm cl}[\Phi]=\left(\frac{m^2}{4\eta^2}\right) \int d^4 \bar{x}\left[\frac{1}{2}(\bar{\partial}_\mu \Phi)^2+\frac{1}{2}\Phi^2-\frac{1}{2}\Phi^3+\frac{\alpha}{8}\Phi^4\right]\quad .
\label{action2}
\eea
The overall factor is determined by the dimensionless parameter 
\bea
\beta=\frac{m^2}{4\eta^2}=\frac{1}{\lambda}\left(1-\frac{\epsilon}{2\lambda a^4}+\dots\right)\quad .
\label{beta}
\eea
For our semiclassical tunneling analysis, we assume $\beta\gg 1$. The quartic coupling strength in (\ref{action2})  is determined by the dimensionless quantity
\bea
\alpha=\frac{\lambda m^2}{4\eta^2}=1-\frac{\epsilon}{2\lambda a^4}+\dots   \quad ,
\label{alpha}
\eea
which tends to 1 in the "thin-wall" limit where $\epsilon\to 0$. The dimensionless parameter $\alpha$ determines the shape of the potential, and its deviation from $1$ is a measure of the vacuum energy difference relative to the barrier height. Figure \ref{fig2} shows some plots, for various values of $\alpha$, of the dimensionless potential 
\bea
U(\Phi)=\frac{1}{2}\Phi^2-\frac{1}{2}\Phi^3+\frac{\alpha}{8}\Phi^4\quad .
\label{dimpot}
\eea

The false vacuum decay rate per unit volume and unit time is denoted $\gamma$, and its one-loop expression is \cite{langer,kobzarev,stone,coleman,affleck,voloshin1,weinberg1,voloshin-erice}
\bea
\gamma= \left(\frac{S_{\rm cl}[\phic]}{2\pi}\right)^2 \left|\frac{{\rm det}^\prime \left(-\Box + U^{\prime\prime}(\phic)\right)}{{\rm det}\left(-\Box +U^{\prime\prime}(\Phi_-)\right)}\right|^{-1/2}\, e^{-S_{\rm cl}[\phic]-\delta_{\rm ct}S[\phic]}\quad ,
\label{rate}
\eea
where the prime on the determinant means that the zero modes (corresponding to translational invariance) are removed. Here $\phic$ is a classical solution known as the ``bounce'' solution \cite{coleman}, defined below, and the prefactor terms in (\ref{rate}) correspond to quantum fluctuations about this bounce solution.  The second term in the exponent, $\delta_{\rm ct}S[\phic]$, denotes the counterterms needed for renormalization. The computational challenge is to evaluate the rate $\gamma$ given a particular form of the classical potential. For our particular quartic model this corresponds to computing the rate for various values of $\beta$ and $\alpha$, the dimensionless parameters defined above in (\ref{beta}) and (\ref{alpha}). 

In the language of quantum field theory \cite{stone,coleman,weinberg1}, $\gamma$ is half the imaginary part of the generating functional of the connected Green's functions with the constant external source, $-\frac{\epsilon}{2a}$, divided by the spacetime volume factor, and so we are essentially computing the renormalized effective action for this system \cite{cw,salam}. It is not possible to compute this tunneling rate $\gamma$ analytically. Indeed, it is not even possible to find the classical bounce solution, $\phic$, analytically, let alone the quantum fluctuations about this classical solution. We present here a simple numerical technique to compute $\gamma$ for general forms of $U(\phi)$. This technique involves a combination of an analytical computation and a numerical computation. The analytical part of the computation is related to the regularization and renormalization of the one loop effective action. The numerical part is elementary, and can be implemented straightforwardly and efficiently in Mathematica.

In Section \ref{bounce} we describe how to compute the classical bounce solution numerically with very high precision. In Section \ref{prefactor} we describe how to compute the determinant prefactor arising from quantum fluctuations about the classical bounce solution. In Section \ref{renormalization} we explain how to regularize and renormalize our answer using the angular momentum cutoff regularization and renormalization scheme developed previously in \cite{dunne}. Our approach is related to an elegant technique presented recently by Baacke and Lavrelashvili \cite{baacke}, and in Section \ref{renormalization} we also compare and contrast these two methods. In Section \ref{conclusions} we conclude with some general comments about possible further applications, and in the Appendix we present the derivation of a simple new formula for the contribution to the decay rate $\gamma$ coming from the zero modes.

\section{Computing the Classical Bounce Solution}
\label{bounce}

The first step in computing the false vacuum decay rate $\gamma$ is to find the classical bounce solution, $\phic(r)$, which is an $0(4)$-symmetric stationary point of the classical Euclidean action, with $\phic(r)$ interpolating between the false and true vacuum as $r$ goes from 0 to $\infty$ \cite{coleman,spherical}. The action evaluated on the classical bounce solution determines the leading exponential factor in (\ref{rate}), and the quantum fluctuations about the classical bounce solution lead to the determinant prefactors in (\ref{rate}).

The bounce $\phic(r)$ solves the nonlinear ordinary differential equation
\bea
-\phic^{\prime\prime} -\frac{3}{r}\phic^\prime +\phic-\frac{3}{2}\phic^2+\frac{\alpha}{2} \phic^3=0
\label{bounceeq}
\eea
with boundary conditions
\bea
\phic^\prime(0)&=&0 
\label{bca}\\
\phic(r)&\to& \Phi_-\equiv 0 \quad,\quad {\rm as}\quad r\to\infty\quad .
\label{bcb}
\eea
It is not known how to find $\phic(r)$  analytically in any nontrivial field theory. Much work has been done in the thin-wall approximation, which in practice means finding $\phic$ as an expansion about the point $\alpha=1$, where the two vacua are degenerate. The point of this present paper is to reduce the calculation of the tunneling rate $\gamma$ to a straightforward numerical computation, without relying on any such approximation or expansion. Thus, we begin by  determining $\phic(r)$ numerically. 

To obtain extremely good precision we use a combination of both forward and backward shooting to compute $\phic(r)$. In forward shooting we numerically integrate (\ref{bounceeq}) [using 4th order Runge-Kutta], starting at $r=0$, and we adjust the initial value $\Phi_0\equiv \phic(0)$ until the second boundary condition (\ref{bcb}) is satisfied to a certain level of accuracy. Since one cannot start exactly at $r=0$, we begin at some very small $r=r_0$ and use a Taylor expansion to write 
\bea
\phic(r_0)&=&\Phi_0+\frac{1}{16} r_0^2\left(2\Phi_0-3\Phi_0^2+\alpha \Phi_0^3\right)\nn\\
\phic^\prime(r_0)&=&\frac{1}{8} r_0\left(2\Phi_0-3\Phi_0^2+\alpha \Phi_0^3\right)\quad .
\label{forwardstart}
\eea
\begin{figure}[h]
\includegraphics[scale=0.9]{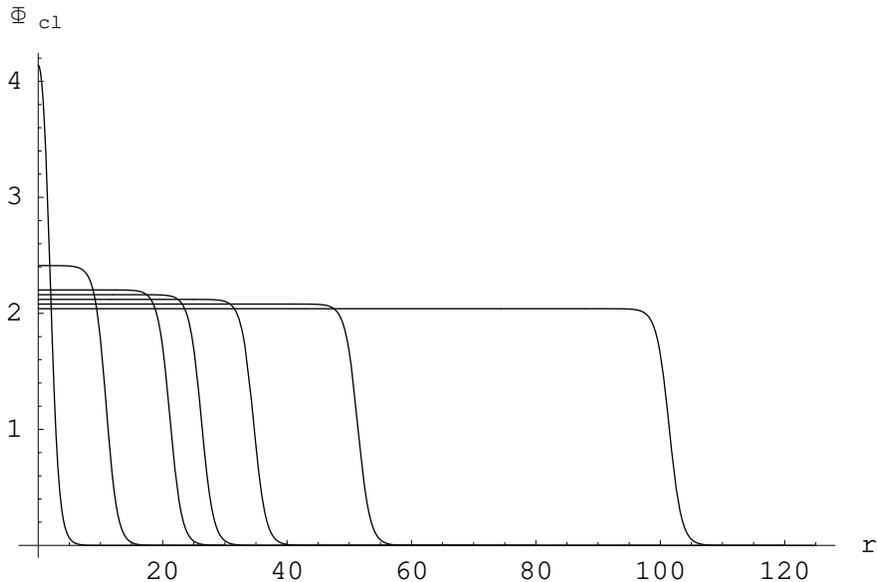}
\caption{Plots of the bounce solution $\phic(r)$ for various values of $\alpha$: $ \alpha= 0.5$, $0.9$, $0.95$, $0.96$, $0.97$, $0.98$, $0.99$, with the plateau ending farther to the right for increasing $\alpha$. Observe that as $\alpha\to 1$, the sharp falloff in $\phic$ occurs at $r\sim \frac{1}{1-\alpha}$, and $\phic$ can be approximated by a step function.}
\label{fig3}
\end{figure}

We adjust the parameter $\Phi_0$ until the large $r$ boundary condition (\ref{bcb}) is satisfied. In backward shooting we begin the numerical integration [also using 4th order Runge-Kutta] at some very large $r=R$, with starting conditions
\bea
\phic(R)=\Phi_\infty\, \frac{K_1(R)}{R}\quad , \quad \phic^\prime(R)=-\Phi_\infty\, \frac{K_2(R)}{R}\quad ,
\label{backwardstart}
\eea
and adjust the parameter $\Phi_\infty$ until the $r=0$ boundary condition (\ref{bca}) is satisfied. 

We first use forward shooting, which produces a good estimate of $\Phi_0$. Then using this bounce solution we estimate $\Phi_\infty$, and use this as a starting point for shooting backwards, which further refines this value of $\Phi_\infty$ and also leads to a refined value of $\Phi_0$. It is a simple matter to obtain 20 - 30 decimal precision for each of $\Phi_\infty$ and $\Phi_0$ in this way.
Some bounce profiles for various values of the coupling parameter $\alpha$ are shown in Figure \ref{fig3}.

\begin{figure}[h]
\includegraphics[scale=0.9]{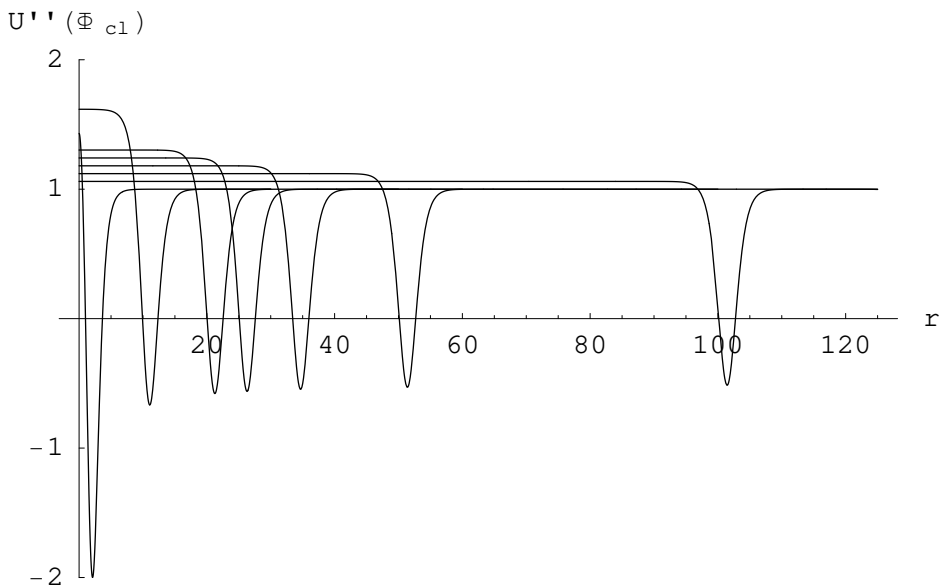}
\caption{Plots of the fluctuation potential $U^{\prime\prime}(\phic(r))$ for various values of $\alpha$ : $\alpha= 0.5$,  $0.9$, $0.95$, $0.96$, $0.97$, $0.98$, $0.99$, with the binding well of the potential appearing farther to the right for increasing $\alpha$. Observe that as $\alpha\to 1$, the potential $U^{\prime\prime}(\phic(r))$ is localized at $r\sim \frac{1}{1-\alpha}$, and is approximated well by the analytic form in \protect{(\ref{thinwall})}.}
\label{fig4}
\end{figure}

Given the bounce solution, $\phic(r)$, the corresponding radial fluctuation potential is 
\bea
U^{\prime\prime}(\phic(r))
= 1-3\phic(r)+\frac{3\alpha}{2}\phic^2(r)\quad .
\label{flucpot}
\eea
Clearly, since $\phic(r)$ is only known numerically, the fluctuation potential $U^{\prime\prime}(\phic(r))$ is also only known numerically.  
Figure \ref{fig4} shows some profiles of this fluctuation potential, corresponding to the various bounce profiles in Figure \ref{fig3}. Note that this fluctuation potential is highly localized, with the localization radius depending strongly on $\alpha$. As $\alpha\to 1$, the fluctuation potential binds states at radius $r\sim \frac{1}{1-\alpha}$, the radius of the bubble of false vacuum. Moreover, in this thin-wall limit, the minimum of the potential approaches $-1/2$, and outside the binding well it approaches 1. In fact,  as noted in Langer's original work \cite{langer}, in this limit $U^{\prime\prime}(\phic(r))$ is approximated well by the analytic potential
\bea
U^{\prime\prime}(\phic(r))\sim 1-\frac{3}{2}\, {\rm sech}^2\left(\frac{1}{2}\left[r-\frac{1}{1-\alpha}\right]\right) \quad .
\label{thinwall}
\eea

\section{Computing the Determinant Prefactor}
\label{prefactor}

Since the bounce solution $\phic(r)$ is a function of $r$, the fluctuation operator $[-\Box +U^{\prime\prime}(\phic)]$ can be decomposed into partial waves, with (dimensionless) radial operators 
\bea
{\mathcal M}_{(l)}=-\frac{d^2}{d r^2}-\frac{3}{r}\frac{d}{d r} +\frac{l(l+2)}{r^2}+1+V(r)\quad ,
\label{radial}
\eea
of degeneracy $(l+1)^2$, with $l=0, 1, 2, \dots$. Here the radial potential $V(r)$ is equal to the fluctuation potential (\ref{flucpot}) with its large radius asymptotic value, $1$, subtracted:
\bea
V(r)= -3\phic(r)+\frac{3\alpha}{2}\phic^2(r)\quad .
\label{radialpot}
\eea
Likewise, the free fluctuation operator $[-\Box +U^{\prime\prime}(\Phi_-)]$ can be decomposed into partial waves, with radial operators 
\bea
{\mathcal M}_{(l)}^{\rm free}=-\frac{d^2}{d r^2}-\frac{3}{r}\frac{d}{d r} +\frac{l(l+2)}{r^2}+1\quad ,
\label{radialfree}
\eea
also of degeneracy $(l+1)^2$.

For each $l$, the ratio of the determinants of ${\mathcal M}_{(l)}$ and ${\mathcal M}_{(l)}^{\rm free}$ can be computed efficiently and precisely using the Gelfand-Yaglom method \cite{gy,levit,erice,forman,kirsten,kirstenbook,kleinertbook}. This result states that for radial operators of the form (\ref{radial},\ref{radialfree}),
\bea
\frac{{\rm det}({\mathcal M}_{(l)})}{{\rm det}({\mathcal M}_{(l)}^{\rm free})}=\left(\lim_{R\to\infty} \left[\frac{\psi_{(l)}(R)}{\psi^{\rm free}_{(l)} (R)}\right]\right)^{(l+1)^2}\quad .
\label{theorem}
\eea
Here  $\psi_{(l)}$ and $\psi_{(l)}^{\rm free}$ are regular solutions of 
\bea
{\mathcal M}_{(l)}\, \psi_{(l)}&=&0\nn\\
{\mathcal M}_{(l)}^{\rm free}\, \psi_{(l)}^{\rm free}&=&0\quad ,
\label{ode}
\eea
with the {\it same} leading behavior at $r=0$. By regularity we can choose this small $r$ behavior to be
\bea
\psi_{(l)}&\sim& r^l \quad , \quad r\to 0 \nn\\
\psi_{(l)}^{\rm free}&\sim& r^l \quad , \quad r\to 0\quad .
\label{r0bc}
\eea
This normalization choice fixes the free solution to be
\bea
\psi_{(l)}^{\rm free}(r)= 2^{l+1} (l+1)! \frac{I_{l+1}(r)}{r}\quad 
\label{freesol}
\eea
where $I_{l+1}(r)$ is the modified Bessel function.

In fact, a numerical improvement comes from realizing that 
both $\psi_{(l)}(r)$ and $\psi^{\rm free}_{(l)} (r)$ grow exponentially fast at large $r$, so it is numerically better to integrate directly the ratio \cite{baacke}
\bea
T_{(l)}(r)\equiv \frac{\psi_{(l)}(r)}{\psi_{(l)}^{\rm free}(r)}\quad ,
\label{ratio}
\eea
which satisfies the equation
\bea
T_{(l)}^{\prime\prime}+\left(\frac{1}{r}+ 2\frac{I_{l+1}^\prime(r)}{I_{l+1}(r)}\right) T_{(l)}^\prime-V(r) T_{(l)}=0\quad ,
\label{tequation}
\eea
with simple {\it initial value boundary conditions}:
\bea
T_{(l)}(0)=1\quad ; \quad T_{(l)}^\prime(0)=0 \quad .
\label{ivbc}
\eea
Then the result (\ref{theorem}) is recast as
\bea
\frac{{\rm det}({\mathcal M}_{(l)})}{{\rm det}({\mathcal M}_{(l)}^{\rm free})}=\left(T_{(l)}(\infty)\right)^{(l+1)^2}\quad .
\label{ttheorem}
\eea

We stress that the result (\ref{ttheorem}) provides a remarkably simple technique for computing the determinant of a radial differential operator. It does not require any detailed knowledge of the spectrum of the operator whose determinant is being computed, nor does it require evaluating and numerically integrating the associated phase shift. Furthermore, the result (\ref{ttheorem}) is ideally suited to numerical evaluation, as initial value boundary conditions are straightforward to implement numerically.

There are three different types of eigenvalue of the fluctuation operator, each having a different role physically and mathematically.

\begin{itemize}

\item
\underline{Negative Eigenvalue Mode : $(l=0)$}

The lowest eigenvalue mode in the $l=0$ sector is a negative eigenvalue mode of the fluctuation operator, and is responsible for the instability leading to decay. It can be identified with homogeneous swelling and shrinking of the bubble of true vacuum. This mode contributes a factor to the decay rate $\gamma$ related to the {\it absolute value}  of the determinant of the $l=0$ fluctuation operator \cite{langer,coleman}. This determinant is computed by numerically integrating (\ref{tequation}) for $l=0$ :
\bea
\left |\frac{{\rm det}{\mathcal M}_{(l=0)}}{{\rm det}{\mathcal M}_{(l=0)}^{\rm free}}\right |^{-1/2}=\left |T_{(0)}(\infty)\right |^{-1/2} \quad .
\label{l=0}\eea

\item
\underline{Zero Eigenvalue Modes : $(l=1)$}

In the $l=1$ sector there is a four-fold degenerate zero eigenvalue of the fluctuation operator. Physically, these four zero modes are the Goldstone modes associated with the breaking of translational invariance. Integrating over the corresponding collective coordinates \cite{gervais} produces the factors of $\frac{S_{\rm cl}}{2\pi}$ in (\ref{rate}). In computing the rate $\gamma$, we need the determinant of the fluctuation operator with the zero modes removed \cite{langer,coleman}. We have found the following simple new formula for the $l=1$ sector prefactor contribution (see Appendix A for a derivation):
\bea
\left(\frac{S_{\rm cl}[\varphi_{\rm cl}]}{2\pi}\right)^2 \left(\frac{{\rm det}^\prime{\mathcal M}_{(l=1)}}{{\rm det}{\mathcal M}_{(l=1)}^{\rm free}}\right)^{-1/2} = \left[\frac{\pi}{2} \Phi_\infty \left(\Phi_0-\frac{3}{2}\Phi_0^2+\frac{\alpha}{2}\Phi_0^3\right)\right]^2\quad ,
\label{zmformula}
\eea
where $\Phi_0=\phic(0)$ is the bounce solution evaluated at the origin, and $\Phi_\infty$ is the coefficient of $K_1(r)/r$ in the large $r$ behavior of $\phic(r)$, as in (\ref{backwardstart}). The advantage of the result (\ref{zmformula}) is that it is expressed entirely in terms of the asymptotic behavior of the classical bounce solution $\phic(r)$. This asymptotic information is already generated in the precise numerical determination of the bounce solution, so no further computation is needed to extract the zero mode contribution to the prefactor.

\begin{figure}[h]
\includegraphics[scale=0.9]{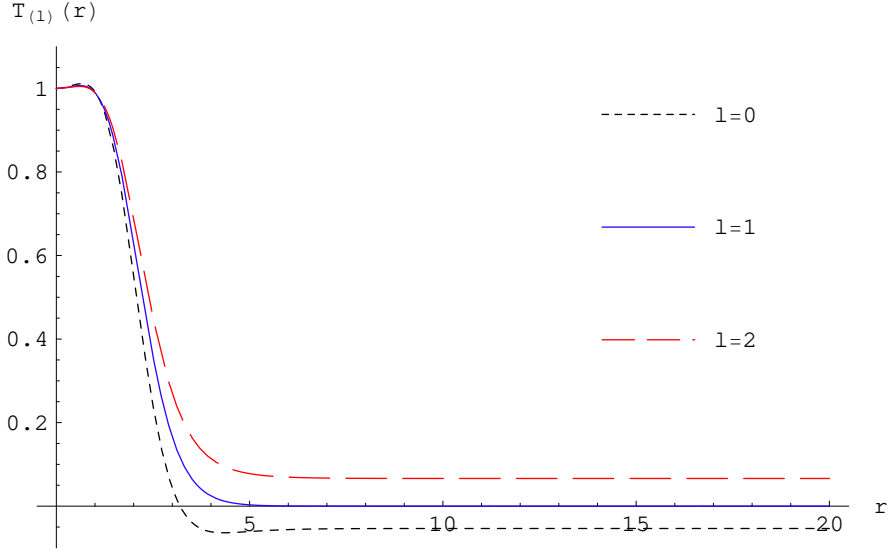}
\caption{Plots of $T_{(l)}(r)$ for $l=0$, $l=1$, and $l=2$. These plots are for $\alpha=0.5$. Note that the asymptotic value, $T_{(l)}(\infty)$, is negative for $l=0$, zero for $l=1$, and positive for $l=2$, confirming the discussion in the text concerning the three different types of modes. }
\label{fig5}
\end{figure}

\item
\underline{Positive Eigenvalue Modes : $(l\geq 2)$}

For $l\geq 2$, the fluctuation operator has positive eigenvalues, each of degeneracy $(l+1)^2$. These modes correspond to deformations of the bubble shape and thickness. For each $l$, the associated radial determinant is computed simply by numerical integration of (\ref{tequation}) with the initial value boundary conditions (\ref{ivbc}):
\bea 
\left(\frac{{\rm det}{\mathcal M}_{(l)}}{{\rm det}{\mathcal M}_{(l)}^{\rm free}}\right)^{-1/2} = \left[T_{(l)}(\infty)\right ]^{-(l+1)^2/2} \quad .
\label{lgeq2}
\eea
\end{itemize}

Figure \ref{fig5} shows plots of $T_{(l)}(r)$ as a function of $r$, for $l=0$, $l=1$, and $l=2$. These plots are for $\alpha=0.5$. Observe that $T_{(l)}(r)$ approaches very rapidly its asymptotic value, $T_{(l)}(\infty)$, for $r$ outside the range of the binding well of the fluctuation potential (compare with Figure \ref{fig4} for $\alpha=0.5$). Also note that the asymptotic value, $T_{(l)}(\infty)$, is negative for $l=0$, zero for $l=1$, and positive for $l=2$, confirming the discussion above of the three different types of modes. 

\begin{figure}[ht]
\includegraphics[scale=0.9]{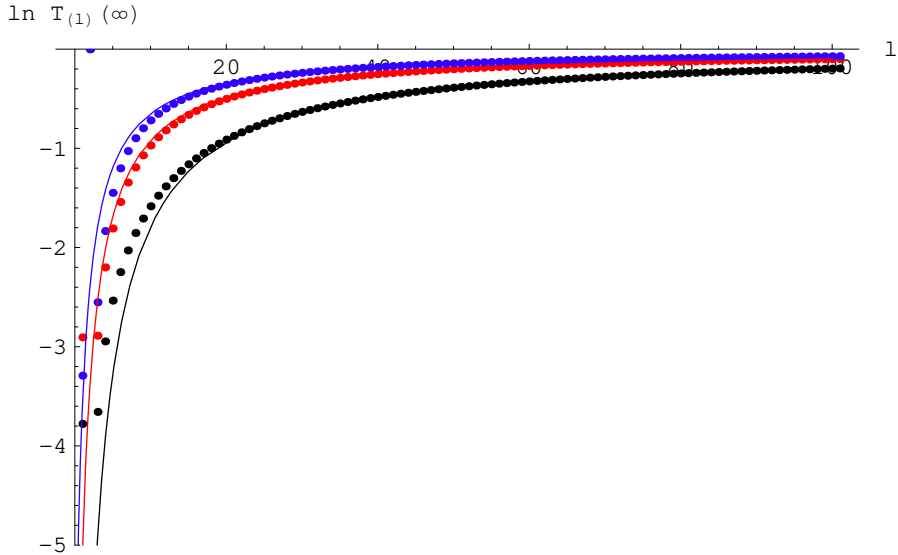}
\caption{The $l$ dependence of $\ln T_{(l)}(\infty)$ for three different values of $\alpha$: $\alpha=0.01$ (top curve), $\alpha=0.7$ (middle curve), and $\alpha=0.9$ (bottom curve). The dots show the values of $\ln T_{(l)}(\infty)$ evaluated numerically using (\ref{tequation}), while the solid lines show the WKB prediction (\ref{largel}) of the leading large $l$ behavior. Notice the excellent agreement for $l\geq 20$.}
\label{fig6}
\end{figure}
For each partial wave with $l\geq 2$, the radial determinant, $\left[T_{(l)}(\infty)\right ]^{(l+1)^2}$, is finite and simple to evaluate.  In the discussion of renormalization it proves convenient to consider the {\it logarithm} of the determinant factors appearing in the rate (\ref{rate}):
\bea 
-\frac{1}{2}\ln \left(\frac{{\rm det}{\mathcal M}_{(l)}}{{\rm det}{\mathcal M}_{(l)}^{\rm free}}\right) = -\frac{1}{2}\left(l+1\right)^2 \ln T_{(l)}(\infty)    \quad .
\label{lnform}
\eea
For large $l$ we can use the radial WKB analysis of \cite{wkb,dunne} to find the leading behavior:
\bea
\ln T_{(l)}(\infty) 
\sim \frac{1}{(l+1)}\left[\frac{1}{2}\int_0^\infty dr\, r \, V(r)\right]+O\left(\frac{1}{(l+1)^3}\right)\quad .
\label{largel}
\eea
The leading large $l$ behavior (\ref{largel}) is illustrated in Figure \ref{fig6} for several different values of the quartic coupling parameter $\alpha$. Notice the very good agreement between the numerical results  for $\ln T_{(l)}(\infty)$ [solid points] and the WKB estimate [curves from (\ref{largel})]. Further subleading corrections to (\ref{largel}) are computed below in the next Section -- see (\ref{wkbnlo}).

\section{Regularization and Renormalization}
\label{renormalization}

The large $l$ behavior (\ref{largel}) means that the formal sum of contributions (\ref{lnform}) to $-\ln \gamma$,
\bea
\frac{1}{2}\sum_{l=2}^\infty (l+1)^2 \ln \left[T_{(l)}(\infty)\right]\quad ,
\label{rawsum}
\eea
is quadratically divergent. This divergence should not be too surprising, as we have neither regulated nor renormalized the determinant prefactor in the expression (\ref{rate}) for $\gamma$. In the language of quantum field theory, we need  to compute the {\it renormalized} one-loop effective action for this interacting scalar field theory \cite{stone,coleman,voloshin1,weinberg1}. Here we will apply the {\it angular momentum cutoff regularization and renormalization technique} developed in \cite{dunne} which was successfully applied to the computation of the quark mass dependence of the fermion determinant in an instanton background in QCD.

\subsection{Regularization}

The first step is to introduce a regulator mass $\mu$ using the proper-time representation of the logarithm :
\bea
\left[\ln  \left(\frac{{\rm det}\,{\mathcal M}_{(l)}}{{\rm det}\, {\mathcal M}_{(l)}^{\rm free}}\right)\right]_{\rm reg}=-(l+1)^2 \int_0^\infty \frac{ds}{s}\, \left(\mu^2 s\right)^\varepsilon\, {\rm tr}\left[e^{-s {\mathcal M}_{(l)}}- e^{-s {\mathcal M}_{(l)}^{\rm free}}\right]\quad ,
\label{mu} 
\eea
where the space-time dimension is extended to $4-2\varepsilon$, and we have explicitly extracted the degeneracy factor $(l+1)^2$ from the trace of the radial operators ${\mathcal M}_{(l)}$ and ${\mathcal M}_{(l)}^{\rm free}$ defined in (\ref{radial}) and (\ref{radialfree}). We then split the sum over $l$ of these regulated terms into two parts:
\bea
\frac{1}{2}\sum_{l=2}^\infty \left[\ln \left(\frac{{\rm det}\,{\mathcal M}_{(l)}}{{\rm det}\,{\mathcal M}_{(l)}^{\rm free}}\right)\right]_{\rm reg}
&=& \frac{1}{2}\sum_{l=0}^L\left[\ln \left(\frac{{\rm det}\,{\mathcal M}_{(l)}}{{\rm det}\,{\mathcal M}_{(l)}^{\rm free}}\right)\right]_{\rm reg}
 +\frac{1}{2}\sum_{l=L+1}^\infty\left[\ln \left(\frac{{\rm det}\,{\mathcal M}_{(l)}}{{\rm det}\,{\mathcal M}_{(l)}^{\rm free}}\right)\right]_{\rm reg}\nn\\\nn\\
 &\equiv& \Sigma_1+\Sigma_2
\label{split}
\eea
Here  $L$ is a large but finite integer (in practice, we take $L$ to be of the order of 50 to 100, depending on the value of $\alpha$). Since $\Sigma_1$ is a finite
sum we can safely remove the regulator $\mu$ and write 
\bea
\Sigma_1= \frac{1}{2} \ln \left |T_{(0)}(\infty)\right | -2 \ln \left[\frac{\pi}{2} \Phi_\infty \left(\Phi_0-\frac{3}{2}\Phi_0^2+\frac{\alpha}{2}\Phi_0^3\right)\right] + 
\frac{1}{2}\sum_{l=2}^L (l+1)^2 \ln T_{(l)}(\infty) 
\label{sum1}
\eea
The first term in $\Sigma_1$ corresponds to the negative mode contribution (\ref{l=0}), the second term is from the zero mode contribution (\ref{zmformula}), and the final sum corresponds to the contributions in partial waves with $2\leq l\leq L$.
Each term $T_{(l)}(\infty)$ in this last sum is evaluated numerically by solving (\ref{tequation}). The second sum in (\ref{split})
\bea
\Sigma_2\equiv -\frac{1}{2}\sum_{l=L+1}^\infty (l+1)^2 \int_0^\infty \frac{ds}{s}\, \left(\mu^2 s\right)^\varepsilon\, {\rm tr}\left[e^{-s {\mathcal M}_{(l)}}- e^{-s {\mathcal M}_{(l)}^{\rm free}}\right] \quad ,
\label{sum2}
\eea
cannot be computed numerically using (\ref{tequation}) because of the infinite sum and the presence of the regulator mass $\mu$. Instead, we use radial WKB, which is a good approximation for large $l$, to compute {\it analytically} the leading large $L$ behavior of  $\Sigma_2$. This is a straightforward computation using the results of \cite{wkb,dunne}. The result is:
\bea
\Sigma_2 &=&\frac{1}{2}\left\{ -\frac{(L+1)(L+2)}{4}\int_0^\infty dr \, r\, V(r)+\frac{\ln L}{8}\int_0^\infty dr\, r^3\, V(V+2)\right.\nn\\
&&\left. -\frac{1}{16} \int_0^\infty dr\,\left(\frac{1}{\varepsilon}+2-\gamma_E+\ln \frac{\mu^2 r^2}{4}\right)\, r^3\, V(V+2)\right\}+0\left(\frac{1}{L}\right) \quad .
\label{wkbresult}
\eea
Here $\gamma_E$ is Euler's constant.
Note that this WKB computation reveals large $L$ divergences going like $L^2$, $L$ and $\ln L$, in addition to a term which is finite in the large $L$ limit. These last two terms are exactly of the form expected for renormalization, as is discussed in the next section. 

The most important observation at this stage is that the large $L$ divergences of $\Sigma_2$ cancel {\it exactly} the large $L$ divergence found numerically in $\Sigma_1$, leaving a finite answer. Indeed, comparing (\ref{wkbresult}) with (\ref{largel}) we see immediately that the quadratic divergence cancels. Extending (\ref{largel}) to include 2nd order WKB contributions proves the cancellation of the sub-leading divergences also.

\subsection{Renormalization}
\label{ren}

In standard perturbative renormalization theory, the self-interacting scalar field theory described
by the action (\ref{action}) involves two renormalization parameters, $\lambda$ and
$a^2$, together with the wave function renormalization $Z$. The renormalization counter
terms can be found by replacing those parameters with $\lambda+ \delta \lambda$, and
$a^2 + \delta a^2$. The one-loop approximations can be found in quantum field theory text books (see, {\it e.g.} \cite{peskin}) :
\bea
\delta \lambda&=&\frac{9 \lambda^2}{32 \pi^2} \left(\frac{1}{\varepsilon} -\gamma_E\right)  \\
\delta (\lambda a^2)&=&\frac{3 \lambda^2 a^2}{32 \pi^2} \left( \frac{1}{\varepsilon}-\gamma_E\right),
\eea
and $Z=1$, in the $\overline{{\rm MS}}$ renormalization scheme using dimensional regularization.
With this replacement we identify the renormalization counterterms as
\bea
\delta_{ct} S&=&\int d^4 x\left[\frac{ \delta \lambda}{8} \left(\phi^2-a^2\right)^2
-\frac{\lambda \delta a^2}{4} \left(\phi^2-a^2\right) \right]   \label{ct1}\\
&=&\frac{1}{32} \int_0^\infty dr \left( \frac{1}{\epsilon}-\gamma_E \right) r^3 V(V+2).
\label{ct2}
\eea
In obtaining (\ref{ct2}) from (\ref{ct1}) we have used  (\ref{expansion}), (\ref{meta}), (\ref{rescale}), (\ref{alpha}), and (\ref{radialpot}) to translate from the original physical field $\phi$, and couplings $\lambda$ and $a^2$, to the field $\Phi$ and coupling $\alpha$ which correspond to expanding the field about the false vacuum $\phi_-$. Notice the appearance in (\ref{ct2}) of the combination $V(V+2)$ when the counterterm is expressed in tems of these fields.
Furthermore, note that the counterterm only involves the pole-term $\frac{1}{\varepsilon}$ and $\gamma_E$,
independent of any other parameters, for instance the mass parameter,
in the  $\overline{{\rm MS}}$ renormalization scheme.  We identify this counterterm (\ref{ct2}) within the WKB result (\ref{wkbresult}), with precisely the right coefficient and structure. We choose this particular  $\overline{{\rm MS}}$ renormalization prescription in order to compare with the work of Baacke and Lavrelashvili \cite{baacke}, who also use an $\overline{{\rm MS}}$ prescription.

\begin{figure}
\includegraphics{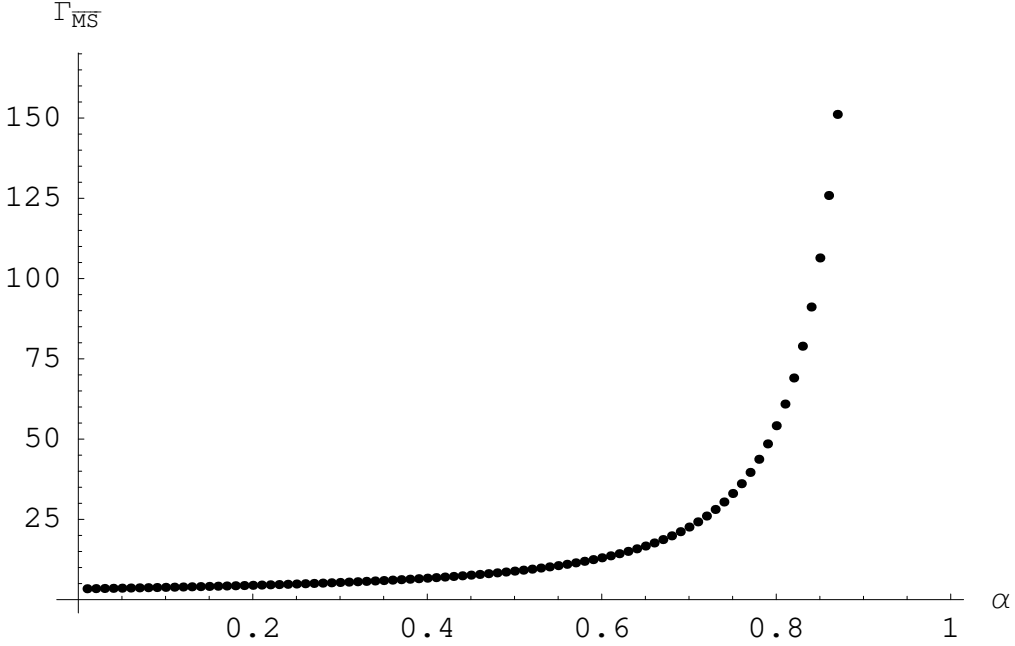}
\caption{Plot of the $\overline{{\rm MS}}$ renormalized effective action (\ref{dm-answer}) as a function of $\alpha$.}
\label{fig7}
\end{figure}

Combining $\Sigma_1$ and $\Sigma_2$ with the above renormalization counter term $\delta_{ct}S$, we get the
one-loop effective action:
\bea
\Gamma_{\overline{{\rm MS}}}&=& \frac{1}{2} \ln \left |T_{(0)}(\infty)\right |-2 \ln \left[\frac{\pi}{2} \Phi_\infty \left(\Phi_0-\frac{3}{2}\Phi_0^2+\frac{\alpha}{2}\Phi_0^3\right)\right]  \nn\\
&& +\frac{1}{2} \lim_{L\to \infty}\left\{\sum_{l=2}^L (l+1)^2 \ln \left(T_{(l)}(\infty)\right)-\frac{(L+1)(L+2)}{4}\int_0^\infty dr \, r\, V(r)\right. \nn\\
&&\left.    +\frac{\ln L}{8}\int_0^\infty dr\, r^3\, V(V+2)\right\}-\frac{1}{16} \int_0^\infty dr\,\left(1+\ln \frac{r}{2}\right)\, r^3\, V(V+2)
\label{dm-answer}
\eea
Here, the first two terms correspond to the $l=0$ contribution (for the negative mode), computed using (\ref{l=0}), and the $l=1$ contribution (for the zero modes), computed using (\ref{zmformula}).
The renormalized effective action $\Gamma_{\overline{{\rm MS}}}$ is finite, and we obtain excellent convergence (accelerated by third order Richardson extrapolation \cite{bender}) for $L$ of the order of 50 to 100, depending on the value of $\alpha$. This result for $\Gamma_{\overline{{\rm MS}}}$ is plotted in Figure \ref{fig7} as a function of $\alpha$.

We can alternatively express (\ref{dm-answer}) with the $L$ dependent WKB subtraction terms included as subtractions {\it inside} the $l$ sum :
\bea
\Gamma_{\overline{{\rm MS}}}&=& \frac{1}{2} \ln \left |T_{(0)}(\infty)\right |-2 \ln \left[\frac{\pi}{2} \Phi_\infty \left(\Phi_0-\frac{3}{2}\Phi_0^2+\frac{\alpha}{2}\Phi_0^3\right)\right]  \nn\\
&& +\frac{1}{2} \sum_{l=2}^\infty (l+1)^2 \left\{ \ln \left(T_{(l)}(\infty)\right)-\frac{\frac{1}{2}\int_0^\infty dr \, r\, V(r)}{(l+1)}+\frac{\frac{1}{8}\int_0^\infty dr\, r^3\, V(V+2)}{(l+1)^3}\right\} \nn\\
&&  -\frac{3}{4}\int_0^\infty dr \, r\, V(r)  +\frac{1}{16} \int_0^\infty dr\, r^3\, V(V+2)\,\left(\frac{1}{2}-\gamma_E-\ln \frac{r}{2}\right)
\label{dm-answer2}
\eea
Here we have simply used 
\bea
\sum_{l=2}^L (l+1)=\frac{1}{2}(L+1)(L+2)-3 \quad ;\quad \sum_{l=2}^L \frac{1}{(l+1)}\sim \ln L-\gamma_E
\label{sums}
\eea
Thus, (\ref{dm-answer2}) shows that we can extend to next-to-leading order the large $l$ behavior  in (\ref{largel}) of $\ln \left(T_{(l)}(\infty)\right)$, the logarithm of the determinant of the radial operators in (\ref{radial}) :
\bea
 \ln \left(T_{(l)}(\infty)\right)\sim \frac{\frac{1}{2}\int_0^\infty dr \, r\, V}{(l+1)}-\frac{\frac{1}{8}\int_0^\infty dr\, r^3\, V(V+2)}{(l+1)^3}+O\left(\frac{1}{(l+1)^5}\right) \quad .
 \label{wkbnlo}
 \eea
We have confirmed numerically that the remainder is indeed $O\left(\frac{1}{(l+1)^5}\right)$. Note that (\ref{wkbnlo}) provides a simple closed-form expression for the large $l$ behavior of $\ln \left(T_{(l)}(\infty)\right)$, and this expression is {\it local} in the fluctuation potential $V(r)$.

It is instructive to compare (\ref{dm-answer2}) with an expression obtained by Baacke and Lavreshavili \cite{baacke} for the same quantity (note that the collective coordinate term, $-2\ln \left(\frac{S_{\rm cl}}{2\pi}\right)$, is separated out in \cite{baacke}), also using the $\overline{{\rm MS}}$ renormalization scheme :
\bea
\Gamma_{BL}&=& \frac{1}{2} \ln \left |T_{(0)}(\infty)\right | +4 \frac{1}{2} \ln T_{(1)}^\prime(\infty)+\frac{1}{2}\left(A^{(1)}_{\rm fin}-\frac{1}{2}A^{(2)}_{\rm fin}\right) \nn\\
&& +\frac{1}{2} \sum_{l=2}^\infty (l+1)^2 \left\{ \ln \left(T_{(l)}(\infty)\right)-h_l^{(1)}(\infty)-\left[h_l^{(2)}(\infty)-\frac{1}{2}\left(h_l^{(1)}(\infty)\right)^2\right]\right\} 
\label{bl-answer}
\eea
The second term corresponds to the fourfold degenerate zero modes, and is computed in \cite{baacke} by evaluating the slope at zero with an additional parameter added to the operator in order to displace the determinant from zero. This agrees numerically with our expression in (\ref{zmformula}), [allowing for the collective coordinate term, $-2\ln \left(\frac{S_{\rm cl}}{2\pi}\right)$], but is more cumbersome to evaluate. The subtractions $h_l^{(1)}(\infty)$ and $h_l^{(2)}(\infty)$ in (\ref{bl-answer}) are associated with first and second order Feynman diagrams \cite{baacke}, and are given explicitly as the asymptotic values of $h_l^{(1)}(r)$ and $h_l^{(2)}(r)$, satisfying the differential equations
\bea
h_l^{(1)\,\prime\prime}+\left(\frac{1}{r}+ 2\frac{I_{l+1}^\prime(r)}{I_{l+1}(r)}\right) h_l^{(1)\,\prime}-V(r)&=&0 \nn\\
h_l^{(2)\,\prime\prime}+\left(\frac{1}{r}+ 2\frac{I_{l+1}^\prime(r)}{I_{l+1}(r)}\right) h_l^{(2)\,\prime}-V(r) h_l^{(1)}&=&0 \quad ,
\label{hequation}
\eea
with {\it initial value boundary conditions}: $h_l(0)=0$, and $h_l^\prime(0)=0$. The remaining terms in (\ref{bl-answer}) are Born approximation terms
\bea
A^{(1)}_{\rm fin}&=&-\frac{1}{8}\int_0^\infty dr\, r^3\, V(r)\nn\\
A^{(2)}_{\rm fin}&=&\frac{1}{128\pi^4}\int_0^\infty dq\, q^3\,\left| \tilde{V}(q)\right|^2 \left(2-\sqrt{q^2+4}\,\ln\left[\frac{\sqrt{q^2+4}+q}{\sqrt{q^2+4}-q}\right]\right)\quad ,
\label{bla}
\eea
where $\tilde{V}(q)$ is the Fourier transform of the fluctuation potential $V(r)$.

Note that Baacke and Lavrelashvili's expression (\ref{bl-answer}) also gives a {\it finite} answer for the renormalized effective action. This means that we have another, quite different, expression for the large $l$ behavior of $\ln \left(T_{(l)}(\infty)\right)$:
\bea
\ln \left(T_{(l)}(\infty)\right)\sim h_l^{(1)}(\infty)+\left[h_l^{(2)}(\infty)-\frac{1}{2}\left(h_l^{(1)}(\infty)\right)^2\right]+\dots 
\label{bl-largel}
\eea
Comparing with (\ref{wkbnlo}), we deduce that the difference between these large $l$ behaviors should agree to order $O\left(\frac{1}{(l+1)^4}\right)$. We have verified this numerically, and in fact we find that  there is a nonzero difference at $O\left(\frac{1}{(l+1)^5}\right)$ :
\bea
&&\left\{h_l^{(1)}(\infty)+\left[h_l^{(2)}(\infty)-\frac{1}{2}\left(h_l^{(1)}(\infty)\right)^2\right]\right\}\nn\\
&&\qquad\qquad  -\left\{\frac{\frac{1}{2}\int_0^\infty dr \, r\, V(r)}{(l+1)}-\frac{\frac{1}{8}\int_0^\infty dr\, r^3\, V(V+2)}{(l+1)^3}\right\}\sim O\left(\frac{1}{(l+1)^5}\right) 
\label{diff}
\eea

Similarly, comparing the terms {\it outside} the $l$ sums in (\ref{dm-answer2}) and (\ref{bl-answer}), we see that these terms in (\ref{dm-answer2}) are {\it local} in the fluctuation potential $V(r)$, while $A^{(2)}_{\rm fin}$ in (\ref{bl-answer}) is {\it nonlocal} in $V(r)$. Nevertheless, the total quantities $\Gamma_{\overline{MS}}$ and $\Gamma_{BL}$ are {\it equal} (allowing for the collective coordinate term, $-2\ln \left(\frac{S_{\rm cl}}{2\pi}\right)$, that is separated out in \cite{baacke}), and we have confirmed this equality numerically. This means that (\ref{dm-answer2}) and (\ref{bl-answer}) are different ways of regulating the $l$ sum, and since each is renormalized using $\overline{MS}$, they should indeed be equal. In other words, in (\ref{dm-answer2}) and (\ref{bl-answer}), the different subtractions inside the $l$ sums are compensated for by different finite pieces outside the $l$ sums, in such a way that the total agrees. This is very similar to behavior found in the computation of mass shifts in soliton theories (see Appendix B of \cite{jaffe}). It is also a highly nontrivial check on both expressions. But, although the expressions (\ref{dm-answer2}) and (\ref{bl-answer}) are equal, in purely pragmatic computational terms the local expression (\ref{dm-answer2}) is considerably simpler. First, the zero mode part is more easily and more accurately computed using (\ref{zmformula}), since it only requires the asymptotic properties of the (already computed) bounce solution $\phic(r)$. Second, one does not need to solve the differential equations (\ref{hequation}) to find $h_l^{(1)}(\infty)$ and $h_l^{(2)}(\infty)$; instead, the large $l$ behavior of $\ln T_{(l)}(\infty)$ is given by local expressions in $V(r)$. Third, one does not need to compute (numerically) the Fourier transform of the fluctuation potential, nor integrate the result over all momentum, as is necessary to compute $A_2$ in (\ref{bla}). Instead, all subtraction terms in (\ref{dm-answer2}) are purely {\it local} in the fluctuation potential $V(r)$.

In a series of papers studying the thin-wall approximation, Konoplich and Rubin \cite{konoplich,kr} have computed this same quantity with a related $\overline{\rm MS}$ renormalization prescription, and obtained the following result in the $\alpha\to 1$  limit [we have translated their notation to match ours]:
\bea
\Gamma_{\rm KR}\sim \frac{9}{32}\, \left[1-\frac{2\pi}{9\sqrt{3}}\right]\,\frac{1}{\left(1-\alpha\right)^3}  +O\left(\frac{1}{1-\alpha}\right) \quad .
\label{kr-answer}
\eea
\begin{figure}[ht]
\includegraphics{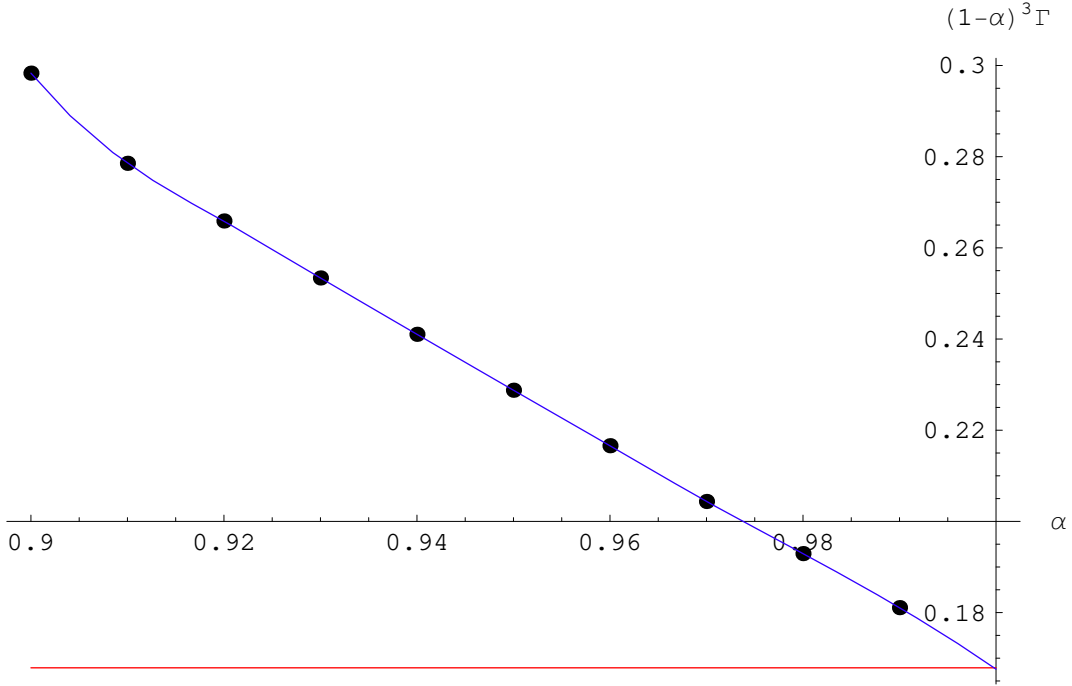}
\caption{Plot of the $\overline{{\rm MS}}$ renormalized effective action, $\Gamma_{\overline{{\rm MS}}}$, in (\ref{dm-answer}), rescaled by  a factor of $(1-\alpha)^3$, as a function of $\alpha$ in the thin-wall approximation limit as $\alpha\to 1$. The solid dots denote our data points for $\alpha$ in steps of $0.01$ from $0.9$ to $0.99$, and the solid  curve is an interpolation through these points. The horizontal line is the leading term of the Konoplich-Rubin result (\ref{kr-answer}), also rescaled by a factor of $(1-\alpha)^3$. Note the excellent agreement at $\alpha=1$.}
\label{fig8}
\end{figure}
Their renormalization prescription is such that this leading term can be compared directly to ours. The subleading terms in (\ref{kr-answer}) are scheme dependent, except for a universal logarithmic term \cite{konoplich,kr}. The agreement with our result (\ref{dm-answer2}) is striking.
Because of the divergence as $\alpha\to 1$, for plotting purposes it is useful to extract the overall factor of $\frac{1}{(1-\alpha)^3}$. In Figure \ref{fig8} we plot our result (\ref{dm-answer2}), rescaled by a factor of $(1-\alpha)^3$, compared with the Konoplich-Rubin thin-wall approximation answer (\ref{kr-answer}), also rescaled by a factor of $(1-\alpha)^3$. The intercept at $\alpha=1$ is given by (\ref{kr-answer}) to be
\bea
\frac{9}{32}\, \left[1-\frac{2\pi}{9\sqrt{3}}\right]\approx 0.16788888 \quad,
\label{intercept}
\eea
which matches perfectly the $\alpha\to 1$ limit of our result (\ref{dm-answer2}), as can be clearly seen in Figure \ref{fig8}.

The $\overline{{\rm MS}}$ prescription provides simple renormalization structures, 
but we can also use a ``physical'' renormalization prescription. For example, we may impose the following normalization condition to the one-loop effective potential,
\bea
\frac{d^2}{d \phi^2} \left(U_{\rm cl}+U_{\rm eff}\right)|_{\phi=\phi_{1}}= \frac{\lambda }{2}(3\phi_1^2-a^2), 
\qquad  \frac{d^4}{d \phi^4} \left(U_{\rm cl}+U_{\rm eff}\right)|_{\phi=\phi_{1}}= 3\lambda.
\label{normal}
\eea
In (\ref{normal}), $ \phi_1\equiv\phi_{-}+\delta \phi_{-} $ is the solution of the one-loop modified field equation
\bea
\frac{d}{d \phi} \left(U_{\rm cl}+U_{\rm eff}\right)|_{\phi=\phi_1}=0. 
\eea 
These normalization conditions can be satisfied when we change the renormalized parameters by finite 
amounts as $ \lambda+\delta\lambda \to \lambda+\delta\lambda+\delta_{fin}\lambda$, and
$ \lambda a^2 +\delta(\lambda a^2) \to \lambda a^2 +\delta(\lambda a^2)+\delta_{fin}(\lambda a^2)$.

\section{Conclusions}
\label{conclusions}

In conclusion, we have presented a simple and practical technique for evaluating the prefactor determinant in the expression for the metastable decay rate in scalar field theories. The technique for computing the determinant is based on the Gelfand-Yaglom formula \cite{gy} for computing the determinant of a one-dimensional (here, radial) differential operator in terms of the asymptotic boundary value of an associated differential equation with initial value boundary conditions. This technique is extremely easy to implement for any given partial wave, but the naive sum over partial waves is divergent. Thus the direct application of the Gelfand-Yaglom formula to higher dimensions is not possible. However, this divergence can be regulated in various ways. Here we propose using the {\it angular momentum cutoff} regularization and renormalization scheme, which has been used previously to compute the explicit mass dependence of the fermion determinant in QCD for massive quarks in an instanton background \cite{dunne}. In this approach the contribution of the low partial waves is computed numerically using the Gelfand-Yaglom formula, while the contribution of the high partial waves is computed analytically using radial WKB, which is a good approximation for large $l$. The merging of these two parts involves renormalization and we have illustrated our technique using an  $\overline{{\rm MS}}$ scheme in order to compare with the work of Baacke and Lavrelashvili \cite{baacke}, who also use the Gelfand-Yaglom technique, also with an $\overline{{\rm MS}}$ prescription, but with a different regularization technique. We have shown that the two techniques agree, and have argued that the one presented here is somewhat easier to implement as it is purely {\it local} in the numerically determined fluctuation potential. The conversion to other renormalization shemes can be done using conventional field theory techniques, and corresponds to including finite polynomial terms in the effective potential. In an appendix we have derived a simple new formula for the $l=1$ determinant with the zero modes removed, solely in terms of the asymptotic values of the bounce solution.

The goal of this work has been to reduce the computation of field theoretic  one loop fluctuation determinants to a straightforward numerical exercise. As a by-product, we have found in (\ref{dm-answer2}) a simple extension of the Gelfand-Yaglom result to higher dimensional radially separable problems. 
There are many possible further applications of this technique, as there are many semiclassical problems where the classical solution, about which one is computing the quantum fluctuations, has radial symmetry. Closely related possible applications include: (i) metastable decay in theories with more than one field \cite{kusenko}, where analytic and approximate approaches to the prefactor are quite difficult, but a direct numerical approach might be more useful; (ii) models in dimensions other than 4, which have been studied in and beyond the thin-wall approximation \cite{voloshin1,nicole,munster,voloshin2+1}; (iii) an extension to finite temperature, where the high temperature limit is essentially a dimensionally reduced 3d radial problem \cite{linde,affleck2}, but for intermediate temperatures the explicit summation over Matsubara modes is necessary.
This technique for computing precisely the fluctuation contribution may also be useful for a set of fascinating questions concerning the validity of Langer's homogeneous nucleation picture itself, as well as the semiclassical approximation \cite{strumia2}. More generally, our technique for extending the Gelfand-Yaglom formula to higher dimensions can also be applied to other symmetric semiclassical configurations such as vortices and monopoles (instantons were considered already in \cite{dunne}). 

\vskip .5cm
{\bf Acknowledgments:}  We thank Holger Gies for helpful discussions. GD thanks the US DOE for support through the grant DE-FG02-92ER40716, and gratefully acknowledges the hospitality and support of Choonkyu Lee and the Center for Theoretical Physics at Seoul National University where part of this work was done.

\section{Appendix A : Zero Mode Contribution}
\label{zeromode}

In this Appendix we present a derivation of the expression (\ref{zmformula}) for the factor contributed by the $l=1$ zero modes. We adapt a method of McKane and Tarlie \cite{mckanetarlie} (see also \cite{kirsten,kirstenbook,kleinert}). First, add a small quantity, $k^2$, to the operator possessing the zero mode, and to the corresponding free operator (although this latter addition is not important in the end). Then for small $k^2$ the existence of the four zero modes for ${\mathcal M}$ [we suppress the $(l=1)$ subscript] implies that
\bea
\frac{{\rm det}({\mathcal M}+k^2)}{{\rm det}({\mathcal M}^{\rm free}+k^2)}\sim (k^2)^4 \, \frac{{\rm det}^\prime{\mathcal M}}{{\rm det}\,{\mathcal M}^{\rm free}}\quad , \quad k^2\to 0 \quad .
\label{addk}
\eea
The result (\ref{theorem}) means we can evaluate the LHS of (\ref{addk}) for arbitrary but small $k^2$ as
\bea
\frac{{\rm det}({\mathcal M}+k^2)}{{\rm det}({\mathcal M}^{\rm free}+k^2)}=\left(\lim_{R\to\infty} \left(\frac{\psi_{k^2}(R)}{\psi_{k^2}^{\rm free} (R)}\right)\right)^4\quad ,
\label{zmratio}
\eea
where $\psi_{k^2}$ satisfies
\bea
-\psi_{k^2}^{\prime\prime}-\frac{3}{r}\psi_{k^2}^\prime +\left(\frac{3}{r^2}+1+k^2+V(r)\right)\psi_{k^2}&=&0\quad ,\nn\\
\psi_{k^2}(0)=0\quad ; \quad \psi^\prime_{k^2}(0)=1\, . &&
\label{psik}
\eea
Note that for $l=1$ we can express the $r\to 0$ boundary conditions (\ref{r0bc}) in this initial value form.
The function $\psi_{k^2}^{\rm free}$ satisfies the same equation and boundary conditions as in (\ref{psik}), but with the potential $V$ set to 0. So, one possible approach, as suggested in \cite{baacke}, to computing ${\rm det}^\prime$ is to compute the derivative of $\lim_{R\to\infty} \left(\frac{\psi_{k^2}(R)}{\psi_{k^2}^{\rm free} (R)}\right)$  as $k^2\to 0$. But a more direct and accurate method is as follows.

Define the function $\psi_0(r)$ to be the solution of (\ref{psik}) with $k^2=0$. Then by elementary integration by parts it follows that
\bea
0&=&\int_0^R dr\, r^3 \psi_0 \left\{-\frac{1}{r^3}\left(r^3 \psi_{k^2}^\prime\right)^\prime +\left(\frac{3}{r^2}+1+k^2+V(r)\right)\psi_{k^2}\right\}\nn\\
&=&\left[r^3\left(\psi_0^\prime \psi_{k^2}-\psi_0 \psi_{k^2}^\prime\right)\right]^R_0 +k^2 \int_0^R dr\, r^3 \psi_0 \psi_{k^2} \quad .
\label{byparts}
\eea
Applying the boundary conditions at $r=0$ we obtain
\bea
\psi_{k^2}(R)\left(1-\left[\frac{\psi_0(R)}{\psi_0^\prime(R)}\right] \left[\frac{\psi_{k^2}^\prime(R)}{\psi_{k^2}(R)}\right]\right)=k^2\frac{ \int_0^R dr\, r^3 \psi_0 \psi_{k^2}}{(-R^3 \psi_0^\prime(R))} \quad .
\label{magic}
\eea
The important observation now is that $\psi_0(r)$ is actually the normalizable zero mode of ${\mathcal M}$, and decreases exponentially  as $e^{-r}$ at large $r$. On the other hand, for arbitrarily small but nonzero $k^2$, the solution $\psi_{k^2}(r)$ {\it increases} exponentially as $e^{r\sqrt{1+k^2}}$ at large $r$. Thus at large $R$ and arbitrarily small but nonzero $k^2$, the identity (\ref{magic}) implies that the leading $k^2$ dependence at small $k^2$ is :
\bea
\psi_{k^2}(R)\sim \frac{1}{2}\,k^2\frac{\int_0^R dr\, r^3 \psi_0^2 }{(- R^3 \psi_0^\prime(R))}\quad , \quad k^2\to 0 \quad .
\label{magic2}
\eea
Then (\ref{zmratio}) leads to the following expression for the determinant with zero mode removed:
\bea
\frac{{\rm det}^\prime{\mathcal M}}{{\rm det}\,{\mathcal M}^{\rm free}}=\left(\frac{\int_0^\infty dr\, r^3 \psi_0^2}{\lim_{R\to\infty} \left[-2 R^3 \psi_0^\prime(R) \psi_0^{\rm free}(R)\right]}\right)^4\, .
\label{zeroremoved}
\eea

To simplify this general expression further, recall that the zero mode $\psi_0$ can be expressed in terms of the classical bounce solution :
\bea
\psi_0(r)=\frac{\phic^\prime(r)}{\phic^{\prime\prime}(0)} \quad .
\label{bouncezm}
\eea
This has two important implications. First, the numerator on the RHS of (\ref{zeroremoved}) can be expressed in terms of the classical bounce action :
\bea
\int_0^\infty dr\, r^3 \psi_0^2&=&\frac{1}{2\pi^2 \left(\phic^{\prime\prime}(0)\right)^2} \int d^4 x \left(\phib^\prime\right)^2\nn\\
&=& \frac{1}{2\pi^2 \left(\phic^{\prime\prime}(0)\right)^2}\, 4\, S_{\rm cl}[\phic]
\label{innerproduct}
\eea
Since $\frac{{\rm det}^\prime\,{\mathcal M}}{{\rm det}\,{\mathcal M}^{\rm free}}$ is raised to the power $-1/2$ on the LHS of (\ref{zmformula}), we see that the $S_{\rm cl}[\phic]$ factors cancel.
The second important implication of (\ref{bouncezm}) is that the leading large $R$ behavior of $\psi_0^\prime(R)$ is determined by the coefficient $\Phi_\infty$ in (\ref{backwardstart}). Since $\psi_0^{\rm free}(r)=8 I_2(r)/r$, the denominator in (\ref{zeroremoved}) is
\bea
\left[-2 R^3 \psi_0^\prime(R) \psi_0^{\rm free}(R)\right] &\sim& -2 R^3 \frac{\Phi_\infty}{\phic^{\prime\prime}(0)} \left(\frac{K_1(r)}{r}\right)^{\prime\prime} \bigg|_{r=R} \left(\frac{8 I_2(R)}{R}\right)\nn\\
&\to & -8 \frac{\Phi_\infty}{\phic^{\prime\prime}(0)}\quad , \quad R\to\infty\quad .
\label{zmr}
\eea
Thus we obtain the simple formula
\bea
\left(\frac{S_{\rm cl}[\phic]}{2\pi}\right)^2 \left(\frac{{\rm det}^\prime{\mathcal M}_{(l=1)}}{{\rm det}\,{\mathcal M}_{(l=1)}^{\rm free}}\right)^{-1/2} = \left[2 \pi\,\Phi_\infty \phic^{\prime\prime}(0)\right]^2 \, .
\label{zmf}
\eea
The final result (\ref{zmformula}) follows by noting that $\phic^{\prime\prime}(0)$ may be expressed in terms of $\phic(0)$ using the bounce differential equation (\ref{bounceeq}) and the boundary condition (\ref{bca}).

We stress that the final expression (\ref{zmformula}) only requires knowing the two asymptotic constants, $\Phi_\infty$ and $\phic(0)$, each of which is determined to great precision in the course of finding the classical bounce solution numerically, as described in Section \ref{prefactor}.

\end{document}